\documentclass[twocolumn,showpacs,preprintnumbers,superscriptaddress,amsmath,amssymb]{revtex4}

\usepackage{graphicx}% Include figure files
\usepackage{dcolumn}% Align table columns on decimal point
\usepackage{bm}% bold math
\usepackage{color}

\begin{document}

\title{New Primary Mechanisms for the Synthesis of Rare $^9$Be in Early Supernovae}
\author{Projjwal Banerjee}
\email{projjwal@berkeley.edu}
\affiliation{Department of Physics, University of California, and Lawrence Berkeley National Laboratory,
Berkeley, CA 94720}
\author{Yong-Zhong Qian}
\email{qian@physics.umn.edu}
\affiliation{School of Physics and Astronomy, University of Minnesota, Minneapolis, MN 55455}
\author{W. C. Haxton}
\email{haxton@berkeley.edu}
\affiliation{Department of Physics, University of California, and Lawrence Berkeley National Laboratory,
Berkeley, CA 94720}
\author{Alexander Heger}
\email{alexander.heger@monash.edu}
\affiliation{Monash Centre for Astrophysics, School of Mathematical Sciences,
Monash University, VIC 3800, Australia}
\date{\today}

\begin{abstract}
We present two new primary mechanisms for the synthesis of the rare nucleus $^9$Be,
both triggered by $\nu$-induced production of $^3$H followed by
$^4{\rm He}(^3{\rm H},\gamma)^7{\rm Li}$ in the He shells of core-collapse supernovae.  
For progenitors of $\sim$$8\,M_\odot$, $^7{\rm Li}(^3{\rm H},n_0)^9{\rm Be}$ occurs
during the rapid expansion of the shocked He shell.
Alternatively, for ultra-metal-poor progenitors of $\sim$$11$--$15\,M_\odot$, 
$^7{\rm Li}(n,\gamma)^8{\rm Li}(n,\gamma)^9{\rm Li}(e^-\bar{\nu}_e)^9{\rm Be}$ occurs
with neutrons produced by $^4{\rm He}(\bar{\nu}_e,e^+n)^3{\rm H}$, assuming a hard 
effective $\bar\nu_e$ spectrum from oscillations (which also leads to heavy element 
production through rapid neutron capture) and a weak explosion (so the $^9$Be survives 
shock passage). We discuss the associated production of $^7$Li and $^{11}$B, noting 
patterns in LiBeB production that might distinguish the new mechanisms from others.
\end{abstract}


\preprint{UCB-NPAT-12-020, NT-LBL-12-023}

\maketitle

It was argued four decades ago that interactions between Galactic
cosmic rays (GCRs) and nuclei in the interstellar medium (ISM) could
approximately account for the abundances of $^{6,7}$Li, $^9$Be, and
$^{10,11}$B observed in the present Galaxy \cite{gcr}.
The rarest of these isotopes, $^9$Be has been regarded as special.  
While big bang nucleosynthesis (BBN) produced an initial abundance of $^7$Li
\cite{fields11} and the $\nu$ process in core-collapse supernovae
(CCSNe) may account for much of the Galaxy's current inventory of
$^7$Li and $^{11}$B and some fraction of its $^{10}$B \cite{WHHH}, it
is widely accepted that $^9$Be is produced almost exclusively in the
ISM by GCRs (e.g., \cite{prantzos}). In this regard, high-velocity
ejecta from very energetic CCSNe can be considered similar to GCRs,
though these events may be too rare to generate significant amounts of
$^9$Be in the ISM \cite{fields02}. Recent observations (e.g.,
\cite{boesgaard}) show that there is a linear correlation between
log(Be/H) and log(E/H) with a slope $\sim$$0.9$--$1$ over
$\sim$$3\,$dex, where E stands for O, Mg, Ti, and Fe, all of which
are major primary products of CCSNe. While GCRs could produce 
such a correlation \cite{prantzos}, the data motivated our search for
alternative CCSN mechanisms for primary $^9$Be production. 
Primary mechanisms operating at low metallicities
are potentially observable because their signatures would 
be preserved in local chemical enrichments influenced by just a
few early CCSNe.

Here we describe two new primary mechanisms for the synthesis of
$^9$Be, both occurring in the He shells of early CCSNe and driven by 
interactions of the $\nu$s from the central proto-neutron star (PNS).
We calculate this nucleosynthesis with the hydrodynamic
code KEPLER \cite{kepler} using the most recent models of
ultra-metal-poor massive progenitors evolved with this code.  For
a progenitor of $8.1\,M_\odot$, the He shell, initially at a radius
$r\sim 10^9\,$cm, is exposed to an intense flux of $\nu$s during its
expansion following shock passage.  Production of $^9$Be
occurs through 
$^4{\rm He}(^3{\rm H},\gamma)^7{\rm Li}(^3{\rm H},n_0)^9{\rm Be}$ 
with $^3$H made by $\nu$ reactions on $^4$He.
For progenitors of 11 and $15\,M_\odot$, the reaction 
$^4{\rm He}(\bar\nu_e,e^+n)^3{\rm H}$ in their outer He shells at 
$r\sim10^{10}\,$cm can produce sufficient neutron densities to drive 
both a rapid (\textsl{r}) neutron-capture process, as described recently 
in \cite{bhq}, and a correlated ``mini-\textsl{r} process" in which 
$^7$Li is converted to $^9$Be.
We explore the sensitivities of the $^9$Be yields, as well as the
associated $\nu$-process yields of
$^7$Li and $^{11}$B, to $\nu$ emission spectra, flavor oscillations,
and the explosion energy. We discuss the implications for
observations at low metallicities and consider other mechanisms 
for $^9$Be production at higher metallicities.

In the updated version of KEPLER, a full reaction network is used to
evolve the nuclear composition of a massive star throughout its
lifetime and to follow the nucleosynthesis that accompanies the
explosion.  This includes the nucleosynthesis associated with shock
heating of the star's mantle and $\nu$-process nucleosynthesis
associated with a $\nu$ burst carrying $\sim$$300\,$B (``Bethe'';
$1\,\mathrm{B}= 10^{51}\,$ergs).  All $\nu$ reactions on $^4$He are
included as in \cite{bhq}.  The KEPLER progenitors employed here,
denoted by u8.1, u11, and u15, have initial metallicities (total mass
fraction of elements heavier than He) of $Z=10^{-4}Z_\odot$ and masses
of 8.1, 11, and 15$\,M_\odot$ \cite{heger}.  These models are very
similar to earlier ones evolved with a more limited reaction network
\cite{WHW}.  The three selected stars develop Fe cores by the end of
their evolutions.  We simulate an explosion by driving a piston into
the collapsing progenitor and following the propagation of the
resulting shock wave \cite{kepler}. Consistent with $\nu$ transport calculations 
and $\nu$ signals from SN 1987A, we assume that the PNS cools by $\nu$ 
emission according to $L_\nu(t)=L_\nu (0)\exp(-t/\tau_\nu)$, with an initial
luminosity per species of $L_\nu(0)=16.7\,$B/s and time constant
$\tau_\nu=3\,$s, where $t=0$ marks the launching of the piston.  
The $\nu$ spectra are approximated as Fermi-Dirac distributions with zero
chemical potential and fixed temperatures $T_{\nu_e}$,
$T_{\bar\nu_e}$, and $T_{\nu_x}=T_{\bar{\nu}_x}$ ($x =\mu,\tau$).

A large set of calculations was performed to assess the sensitivity of
the nucleosynthesis to the explosion energy, $\nu$ spectra, and flavor
oscillations. CCSN simulations using progenitors similar to
u8.1 produce weak explosions with energies of $E_{\rm expl}\lesssim
0.1\,$B \cite{kitaura}, while observations suggest $E_{\rm expl}\sim
1\,$B for more massive progenitors of $\sim$$13$--$20\,M_\odot$ (Fig.~1
of \cite{tominaga}).  Below we discuss results for $E_{\rm expl}\sim
0.06$--$0.3\,$B (u8.1) and $0.1$--$1\,$B (u11/u15).
Two sets of $\nu$ temperatures were used:
$(T_{\nu_e},T_{\bar\nu_e},T_{\nu_x})=(4,5.33,8)\,$MeV (H) and
(3,4,6)~MeV (S), which represent the harder and softer spectra
obtained from earlier (e.g., \cite{woosley}) and more recent
\cite{hudepohl} $\nu$ transport calculations, respectively.  For an
inverted $\nu$ mass hierarchy, $\bar\nu_e \leftrightarrow \bar\nu_x$
oscillations occur before $\nu$s reach the He shell, which greatly
increases the rate of $^4{\rm He}(\bar\nu_e,e^+n)^3{\rm H}$ \cite{bhq}. 
We will explore the case of full $\bar\nu_e \leftrightarrow \bar\nu_x$
interconversion.  We label the nucleosynthesis calculations by the
progenitor model, the $\nu$ spectra, and the explosion energy in units
of B, with, e.g., u8.1H.1 indicating progenitor model u8.1, the
harder $\nu$ spectra H, and $E_{\rm expl} = 0.1\,$B.  Calculations
including $\bar\nu_e \leftrightarrow \bar\nu_x$ oscillations are
denoted by a bar above the H or S.  The abundances of $^{28}$Si and
$^{32}$S in the He shells of u11 and u15 are $\sim$$10$--30
times larger than those found in the older models of \cite{WHW}.
Consequently, we also considered modified models u11* and u15* in
which the He-shell abundances of $^{28}$Si and $^{32}$S were reduced
to their former values.  Representative total mass
yields of $^9$Be, $^7$Li, $^{11}$B, and Fe are given in
Table~\ref{tab}.

\begin{table}[h]
%\vspace{-\baselineskip}
\caption{Yields of $^9$Be, $^7$Li, $^{11}$B, and Fe\label{tab} in $M_\odot$}
\begin{ruledtabular}
\begin{tabular}{lr@{}lrrr}
model&\multicolumn{2}{c}{$^9$Be}&\multicolumn{1}{c}{$^7$Li}&
\multicolumn{1}{c}{$^{11}$B}&\multicolumn{1}{c}{Fe}\\
\hline \\[-1.8ex]
u8.1$\overline{\rm H}.06$&$1.57$&$(-10)$&$2.77(-7)$&$1.47(-7)$&$1.89(-3)$\\
u8.1$\overline{\rm H}.1$ &$1.97$&$(-10)$&$2.97(-7)$&$1.35(-7)$&$1.75(-3)$\\
u8.1H.1                  &$1.20$&$(-10)$&$2.47(-7)$&$1.34(-7)$&$1.79(-3)$\\
u8.1$\overline{\rm S}.1$ &$5.02$&$(-11)$&$1.11(-7)$&$5.67(-8)$&$1.79(-3)$\\
u8.1S.1                  &$2.55$&$(-11)$&$8.19(-8)$&$5.05(-8)$&$1.80(-3)$\\
u8.1$\overline{\rm H}.3$ &$2.56$&$(-10)$&$3.03(-7)$&$1.06(-7)$&$1.45(-3)$\\
u11$\overline{\rm H}.1$  &$1.43$&$(-9)$&2.0--2.3$(-7)$&2.2--8.7$(-7)$&$<7.73(-2)$\\
u11*$\overline{\rm H}.1$ &$9.14$&$(-9)$&1.5--1.9$(-7)$&2.6--9.5$(-7)$&$<7.68(-2)$\\
u11*$\overline{\rm H}.3$ &$9.81$&$(-10)$&$3.26(-7)$&$1.09(-6)$&$<8.75(-2)$\\
u15$\overline{\rm H}.1$  &$<5.20$&$(-10)$&$<3.33(-7)$&$<1.34(-6)$&$<4.42(-2)$\\
u15*$\overline{\rm H}.1$ &$<2.92$&$(-9)$&$<3.15(-7)$&$<1.36(-6)$&$<4.42(-2)$\\
u15*$\overline{\rm H}.3$ &$7.21$&$(-10)$&$1.69(-7)$&$9.50(-7)$&$<4.62(-2)$\\
\end{tabular}
\end{ruledtabular}
\vspace{-0.8\baselineskip}
\flushleft
Note: $X(Y) \equiv X\times10^Y$
%\vspace{-0.8\baselineskip}
\end{table}

\begin{figure}[h]
\includegraphics[width=85mm]{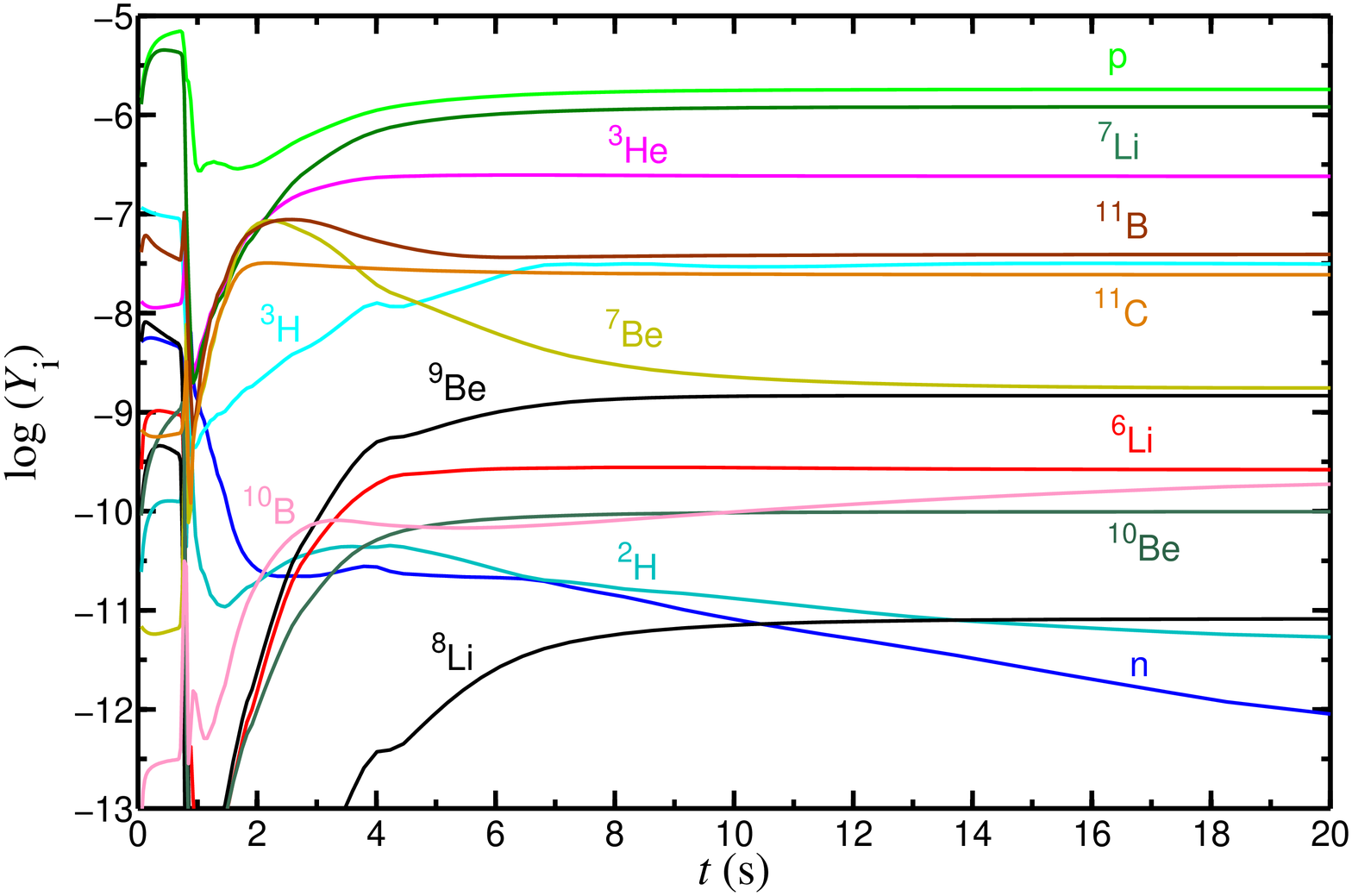}
\caption{Time evolution of the number fraction $Y_i$ for various nuclei in
Zone 95, a typical zone producing $^9$Be in u8.1$\overline{\rm H}.1$.}
\label{fig-u8.1}
\end{figure}

Progenitors of $\sim$$8\,M_\odot$ 
have a steeply-falling density 
profile outside the core and
He-shell radii $r\sim$$10^9\,$cm, in contrast to
$\sim$$10^{10}\,$cm for progenitors of $\gtrsim 11\,M_\odot$. We use Zone 95 in 
u8.1$\overline{\rm H}.1$ to illustrate $^9$Be production in such
progenitors. Prior to shock arrival, the radius, temperature, and density of this zone 
are $1.58\times 10^9\,$cm, $2.21\times 10^8\,$K, and 279~g/cm$^3$, respectively.
The three most abundant nuclei are $^4$He, $^{12}$C, and $^{16}$O with initial
mass fractions of 0.948, 0.043, and 0.009, respectively. The time evolution of the 
number fraction $Y_i$ for various nuclei is shown in Fig.~\ref{fig-u8.1}.
Upon being shocked at $t\sim 0.7\,$s, Zone 95 reaches a peak temperature of 
$\sim$$8\times 10^8\,$K,
so that any $^9$Be produced previously is burned up. 
By $t\sim 5\,$s the 
shocked material has expanded and cooled to $\sim$$2\times 10^8\,$K,
effectively turning off the 
principal destruction reactions $^9$Be$(p,{^4{\rm He}})^6{\rm Li}$ and
$^9$Be$(p,d)2{^4{\rm He}}$.
Yet as the material is still
close to the PNS and the time still relatively early in units of $\tau_\nu$, 
the flux of $\nu$s can efficiently drive the breakup reactions
$^4{\rm He}(\nu,\nu' n)^3{\rm He}$, 
$^4{\rm He}(\nu,\nu' p)^3{\rm H}$, and $^4{\rm He}(\bar\nu_e,e^+ n)^3{\rm H}$.
Production of $^9$Be occurs through
$^4{\rm He}(^3{\rm H},\gamma)^7{\rm Li}(^3{\rm H},n_0)^9{\rm Be}$.
Note that $^9$Be must be produced in the ground state (hence $n_0$) because all of its 
excited states are unstable to breakup. We took the rate
for $^7\mathrm{Li}(^3\mathrm{H},n_0)^9$Be from \cite{brune}.

The  yield of $^9$Be decreases by a factor
of $\sim$4 from u8.1$\overline{\rm H}.1$ to u8.1$\overline{\rm S}.1$ (Table \ref{tab}). 
This reflects the high thresholds of the breakup reactions $^4{\rm He}(\nu,\nu' p)^3{\rm H}$ and 
$^4{\rm He}(\bar\nu_e,e^+ n)^3{\rm H}$, and consequently, their sensitivity to the high-energy
tails of the $\nu$ spectra. The presence of neutral-current production
of $^3$H reduces the sensitivity to $\bar\nu_e \leftrightarrow \bar\nu_x$ oscillations: 
other factors being the same, 
oscillations increase the yield by up to a factor of $\sim$2.
We also explored different explosion
energies based on current CCSN models \cite{kitaura}. 
The change from
u8.1$\overline{\rm H}.1$ to u8.1$\overline{\rm H}.06$ (u8.1$\overline{\rm H}.3$) 
produces a 20\% decrease (30\% increase) in the $^9$Be yield primarily through
the expansion rate of the shocked He-shell material: a higher
$E_{\rm expl}$ results in faster cooling to 
$\sim$$2\times 10^8\,$K and hence, more $^9$Be.
The $^9$Be production in progenitors like u8.1 depends on the 
sharply-falling density structure 
outside the core: the He shell is shocked, expands, and cools on
timescales comparable to $\tau_\nu$, and thus before the $\nu$ flux has diminished significantly.
In such progenitors the yield is essentially independent of the initial
metallicity.

As in the u8.1 models, any $^9$Be produced prior to shock arrival in the inner
He shells of the u11/u15 models is destroyed by shock heating. The subsequent 
$\nu$-induced replenishment of $^3$H and thus $^7$Li that leads to $^9$Be 
production in the rapidly expanding and cooling
He-shell ejecta of the u8.1 models is not replicated in the u11/u15 models, due to the much 
smaller $\nu$ fluxes associated with the latter's much larger He-shell radii and
longer delay in shock arrival there. Instead, the synthesis of $^9$Be in these models 
depends on its preshock buildup in outer He shells and its postshock survival due to
diminished shock heating there.

\begin{figure}[h]
\includegraphics[width=85mm]{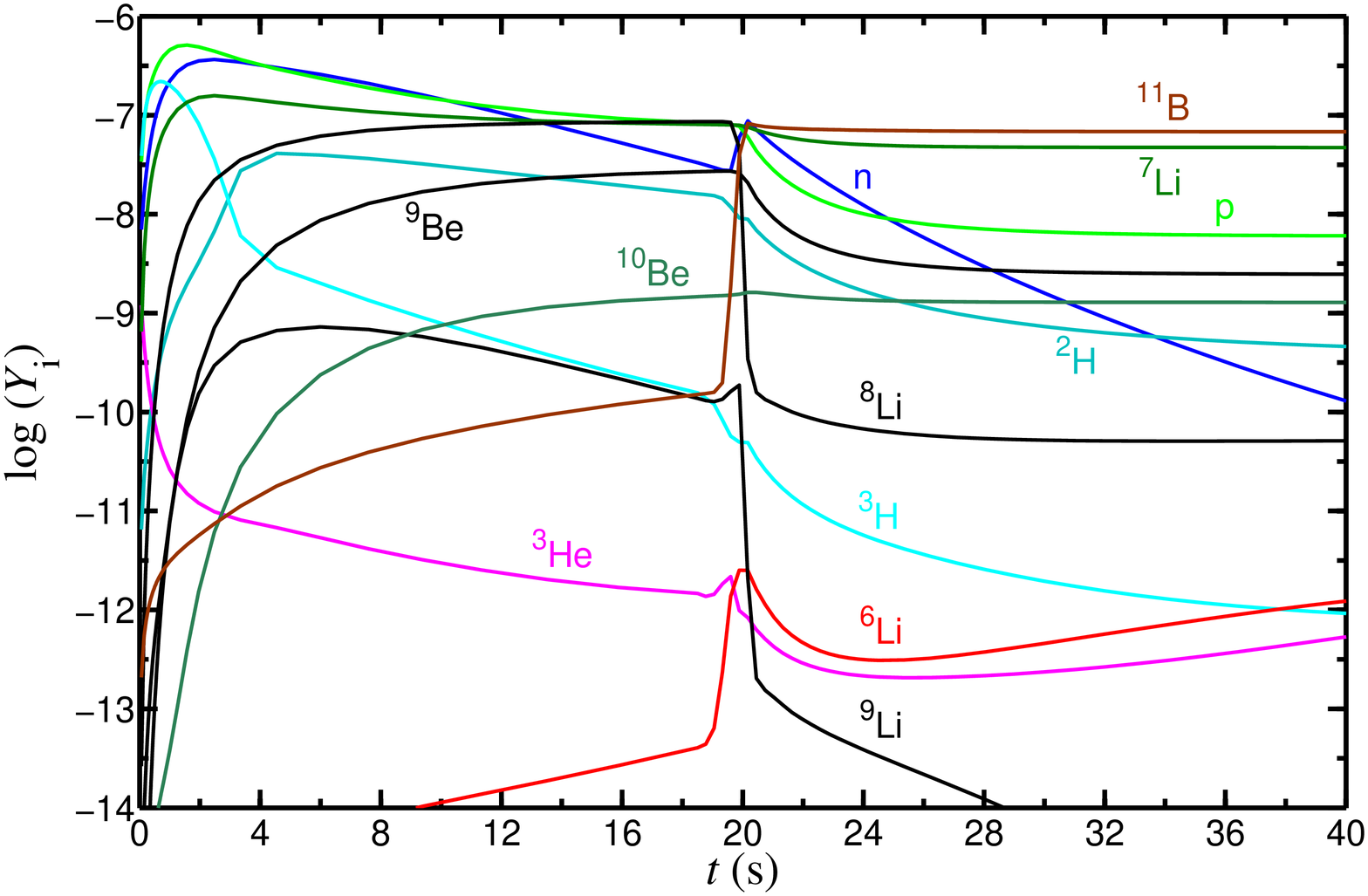}
\caption{Time evolution of the number fraction $Y_i$ for various nuclei in
Zone 223, a typical zone producing $^9$Be in u11*$\overline{\rm H}.1$.}
\label{fig-u11}
\end{figure}

We illustrate this mechanism in u11*$\overline{\rm H}.1$, which is very similar to the model
recently used in \cite{bhq} to demonstrate a $\nu$-induced \textsl{r}-process in outer He shells 
at low metallicities (modifying scenarios discussed earlier in \cite{ech,WHHH}).  As in
\cite{bhq}, we make use of
$^4{\rm He}(\bar\nu_e,e^+ n)^3{\rm H}$ as an important neutron source that can be enhanced by
$\bar\nu_e \leftrightarrow \bar\nu_x$ oscillations.  Zone 223 is representative of $^9$Be production 
in u11*$\overline{\rm H}.1$.
Prior to shock arrival, its radius, temperature, and density
are $1.10\times 10^{10}\,$cm,
$8.49\times 10^7\,$K, and 50~g/cm$^3$, respectively. The composition is nearly pure $^4$He,
with initial mass fractions of
$\sim$$10^{-5}$, $10^{-8}$, $4\times 10^{-9}$, and $4\times 10^{-8}$
for $^{12}$C, $^{28}$Si, $^{32}$S, and $^{56}$Fe, respectively.
Figure~\ref{fig-u11} shows the time evolution of $Y_i$ for important nuclei.

We find that the above neutron source drives both the
\textsl{r}-process described in \cite{bhq} and an
analogous ``mini-\textsl{r} process" through
$^7{\rm Li}(n,\gamma)^8{\rm Li}(n,\gamma)^9{\rm Li}(e^-\bar\nu_e)^9{\rm Be}$.
The $^9$Be yield is limited by the short 838 ms half-life of $^8$Li, which 
$\beta$ decays through the 3.0 MeV resonance in $^8$Be to $^4{\rm He}+{}^4{\rm He}$,
and by the 49.5\% branching ratio for $^9$Li to decay to particle-unstable
excited states in $^9$Be. The ``mini-\textsl{r} process" operates for $\sim$$20\,$s 
prior to shock arrival, at which time the temperature and density jump to 
$\sim$$2\times 10^8\,$K and $\sim$$220\,$g/cm$^3$, respectively, and a
short burst of
neutrons is released by $^8{\rm Li}(^4{\rm He},n)^{11}{\rm B}$ (Fig.~\ref{fig-u11}). 
Incomplete destruction of $^9$Be occurs during the few seconds in which the postshock 
temperature stays at $\sim$$2\times 10^8\,$K, as
protons generated by $^4{\rm He}(\nu,\nu' p)^3{\rm H}$ 
are consumed by $^9{\rm Be}(p,{^4{\rm He}})^6{\rm Li}$ and 
$^9$Be$(p,d)2{^4{\rm He}}$ (Fig.~\ref{fig-u11}).

The production of $^9$Be in the u11/u15 models is sensitive to
the explosion energy (Table~\ref{tab}).  The higher postshock
temperatures found in more energetic explosions greatly enhance $^9$Be
destruction.  For $E_{\rm expl} \gtrsim1\,$B, essentially all of the
$^9$Be made in the He shell prior to shock arrival is destroyed: only
that produced in the oxygen shell through $^{12}{\rm
  C}(\nu,\nu')^3{\rm He}+{}^9 \mathrm{Be}$ \cite{WHHH} survives.  The
absence of protons and $^4$He in the oxygen shell eliminates
light-particle reactions through which $^9$Be can be destroyed.

The $^9$Be yields for the u11/u15 models are also sensitive to the
He-shell composition (Table~\ref{tab}). The abundances of the
hydrostatic-burning products $^{28}$Si and $^{32}$S are
$\sim$$10$--$30$ times higher in u11/u15 than in u11*/u15*.  As
$^{28}$Si and $^{32}$S are neutron sinks, $Y_n$ is consequently
reduced in u11/u15.  The effects on $^9$Be yields are quadratic in
$Y_n$, as two neutrons are captured to produce $^9$Be.  Increasing the
initial metallicity also reduces $Y_n$ due to neutron sinks such as
$^{56}$Fe. He-shell production of $^9$Be dominates oxygen-shell
production only for initial metallicities of
$Z\lesssim10^{-3}Z_\odot$.  A similar bound on He-shell metallicity
comes from requiring a neutron density capable of supporting a
$\nu$-induced \textsl{r}-process \cite{bhq}.  As for this
process, $^9$Be synthesis requires both hard $\nu$ spectra
and $\bar\nu_e \leftrightarrow \bar\nu_x$ oscillations characteristic
of an inverted mass hierarchy so that $\nu$ reactions can produce an
adequate $Y_n$.

In summary, $^9$Be synthesis in the He zones of models like u11/u15
requires an accompanying \textsl{r}-process and $E_{\rm expl}\lesssim
0.3\,$B.  Severe fallback of inner layers was found to occur in all
u11/u15 models with $E_{\rm expl}\lesssim 0.3\,$B.  Table~\ref{tab}
thus only gives ranges or upper limits for the yields when production
partially or entirely occurs in the fallback zones.  The actual yields
in these cases depend on the extent of the poorly-understood mixing
and ejection associated with fallback.

In the models we explored, $^9$Be production is accompanied by
production of $^7$Li and $^{11}$B (Table~\ref{tab}). In all u11/u15
models the latter production operates as described in
\cite{WHHH}, through $\nu$ breakup of $^4$He and $^{12}$C prior
to shock arrival. In contrast, large fractions of $^7$Li and $^{11}$B
are produced in u8.1 models by $\nu$ interactions occurring in the
O-Ne shell that has been severely altered by shock passage.  Because
this shell is so close ($r \sim 1.8\times 10^8\,$cm) to the core, shock
passage leads to complete disassociation of nuclei into free nucleons, which then are
reassembled into $^4$He and Fe-group nuclei as the shell expands and
cools.  The $\nu$ irradiation of this material leads to $^4{\rm
  He}(^3{\rm He},\gamma)^7{\rm Be}(^4{\rm He},\gamma)^{11}{\rm C}$.
The subsequent decay of $^7$Be and $^{11}$C accounts for, e.g.,
$\sim$$70$\% of the $^7$Li and $\sim$$43$\% of the $^{11}$B produced
in u8.1$\overline{\rm H}.1$.  Properties of $^7$Li and $^{11}$B
production in all u8.1 models studied include: 1) weak dependence on
$\nu$ spectra, flavor oscillations, and the explosion energy and no
fallback; 2) stable yields within a factor of $\sim$$2$ of
$1.6\times 10^{-7}$ and $8.6\times 10^{-8}\,M_\odot$, respectively; 3)
a number-yield ratio $^{11}$B/$^7$Li of $\sim$$0.2$--$0.4$, comparable
to the solar value of 0.29 \cite{lodders} and distinct from the values
of $\sim$$1$--$3.6$ found in u11/u15 models. The last point is of
interest because in the more massive progenitors $^{11}$B tends to
overwhelm all other yields \cite{WHHH,austin}.  For all the models in
Table~\ref{tab}, production of $^6$Li and $^{10}$B is negligible with
$^6$Li/$^7\mathrm{Li}\lesssim 10^{-4}$ and
$^{10}$B/$^{11}\mathrm{B}\sim(0.5$--$2)\times10^{-2}$.

Proposed mechanisms for Be production can be tested against observations.
The GCR mechanism (e.g., \cite{prantzos}) is severely constrained by
a recent study \cite{boesgaard} showing that 
[Be/Fe]~$\equiv\log{\rm (Be/Fe)}-\log{\rm (Be/Fe)}_\odot\sim 0\pm0.5$
for a large sample of stars covering 
[Fe/H]~$\sim -3.5$ to $-0.5$. Using the yields in Table~\ref{tab} and the solar
abundances in \cite{lodders}, one obtains
[Be/Fe]~$\sim 0\pm0.2$ for at least three cases:
(I) all the u8.1 models with the hard $\nu$ emission spectra,
(II) u11*$\overline{\rm H}.1$ with the maximum possible Fe yield, and 
(III) u11$\overline{\rm H}.1$, u11*$\overline{\rm H}.3$, and
u15*$\overline{\rm H}.3$
if their actual Fe yields are $\sim$$10\%$ of the upper limits.

At metallicities for which the ISM was enriched mostly by a single CCSN,
yields would be mixed with a total mass of hydrogen of
$\sim$$10^3(E_{\rm expl}/0.1\,\mathrm{B})^{6/7}M_\odot$
\cite{thornton}.  One finds [Fe/H]~$\sim -3.5$ to $-2.8$ (I), $-1.4$
(II), and $-3$ to $-2.4$ (III) for the above cases. The corresponding
enrichment of Li is $A({\rm Li})\equiv\log({\rm Li/H})+12\sim
1$--$1.8$.  This is well below the level due to BBN and thus
consistent with the plateau $A({\rm Li})\sim 2.2$ observed at these
metallicities \cite{fields11}.  Our models also give $A({\rm B})\sim
0.6$--1.3, 1.4--1.9, and 1.3--1.9 with B/Be~$\sim (3$--$9)\times
10^2$, 20--85, and $10^2$--$10^3$ in Cases I, II, and III,
respectively. The limited data on B at [Fe/H]~$\leq-1.4$
show $A({\rm B})\sim 0$--1.8 but typically with rather low B/Be
values of $\sim$$10$--$20$ \cite{btan}. Large B/Be values have been
observed but are usually attributed to greater depletion of Be relative to
B in stars \cite{btan}.  Simultaneous observations of Be and B carried
out for more stars at [Fe/H]~$\leq -1.4$ could test this
interpretation, versus the intrinsic high yield ratio of B/Be for our models. 

While our u8.1 mechanism continues to operate
with increasing metallicity, the mass range of candidate progenitors is
likely to shrink \cite{WHW}, limiting their integrated contribution to the Galactic inventory
of $^9$Be.  Indeed, the low solar value of
B/Be~$\approx 31$ \cite{lodders} requires other $^9$Be production 
mechanisms, e.g., those accompanying the \textsl{r}-process in recent simulations of 
neutron star (NS) mergers \cite{goriely}. Thus we regard both of the new
mechanisms discussed here as low-metallicity, candidate primary
processes that could naturally explain the linear growth of $^9$Be with CCSN-associated
metals at [Fe/H]~$\lesssim -1.4$.

In conclusion, we stress that candidate processes for early $^9$Be production,
including the two primary mechanisms described here as well as adaptations of the GCR
mechanism, elevate the importance of observations to determine both
overall trends in early LiBeB evolution and specific abundance patterns that may characterize
local enrichments in these elements. The more exotic of our two mechanisms requires
special conditions: hard $\nu$ emission spectra, $\bar\nu_e \leftrightarrow
\bar\nu_x$ oscillations, low metallicities, and low explosion energies, the first three of which
are also needed for a simultaneous $\nu$-driven \textsl{r}-process. The
possible correlation of an early \textsl{r}-process with $^9$Be production may be a
feature of these processes that observers could exploit.  Finally, we note that
the nuclear physics of the two mechanisms explored here is not
exotic: the necessary conditions are likely to be found in other astrophysical sites,
with NS \cite{goriely} and white dwarf \cite{shen} mergers being obvious possibilities.

%\begin{acknowledgments}
We thank S. Goriely and H.-T. Janka for sharing their results on
NS mergers. This work was supported in part by the
US DOE [DE-SC00046548 (Berkeley), DE-AC02-98CH10886 (LBL), and
DE-FG02-87ER40328 (UM)];  by the NSF [PHY02-16783 (JINA)];
by ARC Future Fellowship FT120100363 (AH); and by the Alexander von Humboldt
Foundation (WCH).
%\end{acknowledgments}

\end{document}